# The Impact of Photon Entanglement on Local and Nonlocal Dispersion Cancellation

Amin Babazadeh, Rahman Nouroozi[*]

*Physics Department, Institute for Advanced Studies in Basic Sciences, Zanjan, Iran*

[*]*Corresponding author: rahman.nouroozi@iasbs.ac.ir*

**Abstract:**

Recently the engineering of the entanglement for photon pairs generated during the spontaneous parametric down conversion process (SPDC) can be achieved via manipulation of pump wavelength behind a $\chi^{(2)}$-based type II SPDC process [1]. Such effect is used in this paper for demonstration of non-classical dispersion cancellation phenomenon in both local and nonlocal detections, theoretically. The following results are analytically achieved: I) For local detection, if narrow pump laser (highly entangled photons) are used, the dispersive broadening cancelation is directly depends on the degree of entanglement. The higher entanglement degree, the more compensation occurs. The results indicate that by increasing the FWHM of the pump the impact of the entanglement degree is decreased in such a way that for a pump with FWHM=4 nm the entanglement has no effect on the broadening. Therefore, the dispersive broadening is only depends on the temporal walk-off between generated photon pairs and the pump. II) For nonlocal detection, it is also shown that entanglement cancels the dispersive broadening if and only if each of the generated paired photons propagate through dispersive material with identical length with opposite group velocity sign.

## I. Introduction

Entanglement is the base of explanation for many non-classical phenomena e.g. quantum bell states [2], [3], quantum imaging [4], [5] and quantum teleportation [6]–[8]. Also, there are some reports which used entangled photons for dispersion cancellation [9]–[16]. When a pair of entangled photons travels through two different dispersive media the dispersion experienced by one photon can be canceled by the dispersion experienced by the other one. Such a phenomenon is known as dispersion cancellation and proposed initially by J. D. Franson [9] and later on by A. Steinberg [10] in two different scenarios. In Franson's scenario which is illustrated in Fig. 1-a schematically, the proposed concept is a nonlocal-based detection. He reported that if time-frequency entangled photon pairs were sent to different detectors in the presence of two external dispersive medium with equal length and the same group velocity dispersion but with opposite sign, the photon pairs remained tightly coincidence in time. He showed analytically that this dispersion cancellation only was occurred for entangled photon pairs and cannot be described by classical field theory. However its direct experimental demonstration was limited by the temporal resolution of the photo detectors. O'Donnell et al. [17] reported experimental demonstration of the phenomenon by altering nonlocal detection to the local detection. They used the detection of the photons generated via up-conversion of the input entangled photon pairs in a nonlinear crystal. It is important to mention that, such a local detection method is also used for classical light with almost same dispersion cancellation results [18]–[20].

In the another scenario, Steinberg et al. [10], [11] proposed a local measurement of the photons coincidence rate based on Hong-Ou-Mandel (HOM) interference [21] instead of measuring correlation peak (see Fig. 1-b). Since the HOM-based coincidence measurement is an interference effect in a fixed relatively long period of time window, such an experiment which is known as local detection would simplify the demonstration of the dispersion cancellation phenomenon. Therefore the slow response of the detectors does not limit the measurement. In their experiment the time-frequency entangled photon pairs was generated via a continuous wave pumped spontaneous parametric down conversion (SPDC) process. Then the coincidence measurement was utilized in an unbalanced dispersive (e.g. an external dispersion medium in one arm) of the HOM interferometer. The continuous wave pumping causes the uncertainty in the absolute time of photon pair emission and the conclusion confirmed that the HOM interference dip is completely unaffected in the presence of an external dispersion. Later on, J. D. Franson [22] claimed that the second scenario [10] is a local detection experiment and the photon entanglement is not necessary for its demonstration. Franson's argue proved also the classical dispersion cancelation whereby the local detection was performed. These dispersive compensation results which differs from classical dispersive compensation[23], [24] makes an arguments [25], [26] whether entanglement is necessary for dispersion cancellation or not?

It is shown that the degree of entanglement for generated photon pair in a SPDC process can be engineered by tuning of the input pump wavelength [1], [27]. Here in this paper, such a technique is used to manipulate the photon pair entanglement to examine theoretically that the external dispersion cancelation in the local detection scenario is independent of the input photon entanglement for a situation that a broad band pump laser is used. It is shown that for a specific wavelength of the pump in a SPDC process (620 nm with FWHM=4 nm for a periodically polled potassium titanyl phosphate (PPKTP) crystal) the generated photon pairs are not entangled and their Shmidt number (a quantity shows the degree of entanglement and to be explained later on) is equal to one. The entanglement manipulation of the generated photon pairs with different pump wavelengths in the HOM interferometer is studied for the situation that one of the HOM interferometer arm has an external dispersive medium. According to the Franson argue, it is expected that for the wavelength which the states are not entangled the FWHM of the HOM interference should be broaden more than the others. This broadening has not been observed for a pump with 4 nm spectral bandwidth. The results presented in this paper for broadband spectral bandwidth (here 4 nm of pump FWHM) conclude that the temporal walk-offs between generated photon pairs and the pump have the strong effect on the FWHM of HOM interference. However, decreasing the pump spectral bandwidth and using pair photons with different entanglement degrees and looking to the broadening of the HOM interference show that for narrow spectral bandwidth the entanglement compensate the dispersive broadening.

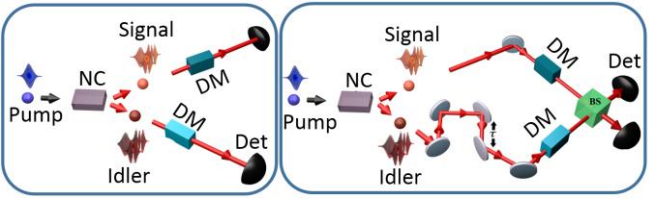

FIG. 1. a) The schematic representation of Franson's nonlocal dispersion cancelation concept; ultrashort pulsed laser used to pump a nonlinear crystal (NC). Entangled photon states with time-frequency Schmidt modes generated during the spontaneous parametric down conversion process. Then each of the paired photons passes through a dispersive material (DM) and detected by detectors (Det). b) The schematic illustration of the Hong-Ou-Mandel interferometer for dispersion cancelation measurement. The photons (signal/idler) can be mixed on a beam splitter (BS) after passing through external dispersive medium. The interval arriving time of idler and signal on the BS can be controlled by adjusting the $\tau$ (e.g. Optical path).

For comparing the local and nonlocal detection scenarios, the same approach is applied for nonlocal detection (Franson's concept) as well. In this case the entanglement shows its effect only for the situation when two dispersive media with identical length but opposite group velocity sign is used.

This paper is organized as follows: In section II the single photon source will be described. In section III and IV the local and nonlocal detection will be explained and the FWHM of dispersive broadening for both detections will be derived analytically. In section V the method of engineering the entanglement degree will be discussed and it is used in section VI and before the conclusions, to illustrate the effect of entanglement degree on the dispersive broadening.

## II. Single photon Source

One promising method to generate a pair of single photons with entangled state is SPDC process in a second order nonlinear ($\chi^{(2)}$) crystal with type II phase-matching. The generated quantum entangled photon states are different spatial-spectral modes if the SPDC is pumped with an ultrashort pulses [28]. Following the standard quantum first order perturbative approach, the state of two photons produced via SPDC process can be written as [29]:

$$|\psi_{SPDC}\rangle \propto \iint d\omega_s d\omega_i \alpha(\omega_s+\omega_i) \times \phi(\omega_s,\omega_i) D(\omega_s,\omega_i) a_s^\dagger(\omega_s) a_i^\dagger(\omega_i)|0\rangle \quad (1)$$

Where $D(\omega_s,\omega_i)$ represents the external material dispersion function which acts as a phase term in the state of the generated signal/idler photons and can be expressed as $D(\omega_s,\omega_i) = exp(i(\beta_s\upsilon_s^2 + \beta_i\upsilon_i^2))$ with group velocity dispersion of $\beta_\mu = l_\mu \partial^2 k_\mu / 2\partial \omega_\mu^2$ along the propagation length $l_\mu$ ($\mu=s,i$) [29]. $a_s^\dagger(\omega_s)$ and $a_i^\dagger(\omega_i)$ denote the creation operators for signal and idler photons, respectively. The pump spectral envelope and the phase matching functions, liable for energy and momentum conservations, are represented by $\alpha(\omega_s+\omega_i)$ and $\phi(\omega_s,\omega_i)$, respectively.

For pump envelope function ($\alpha(\omega_s+\omega_i)$) a Gaussian pump spectrum with a spectral bandwidth of $\sigma_p$ is assumed:

$$\alpha(\nu_s+\nu_i) = \exp\left(-\frac{(\nu_s+\nu_i)^2}{2\sigma_p^2}\right) \quad (2)$$

The frequency detuning, $\nu_\mu = \omega_\mu - \omega_\mu^0$ ($\mu=i,s$) is defined from the perfectly phase matching central frequency $\omega_\mu^0$. The phase matching function ($\phi(\omega_s,\omega_i)$) describing the optical properties of $\chi^{(2)}$ crystal can be written as [30]:

$$\phi(\omega_s,\omega_i) = \left(\exp\left(-\gamma\left(\frac{L}{2}\Delta k(\omega_s,\omega_i)\right)^2\right)\right) \times \left(\exp\left(i\frac{L}{2}\Delta k(\omega_s,\omega_i)\right)\right) \quad (3)$$

Where, $L$ is the length of the nonlinear crystal and $\Delta k(\omega_s,\omega_i) = k_p(\omega_s+\omega_i) - k_s(\omega_s) - k_i(\omega_i)$ is the phase mismatch between pump, signal and idler. In this equation, the phase mismatched *sinc* function has been approximated with a Gaussian profile with $\gamma = 0.193$. By choosing such a value for $\gamma$, same FWHM can be achieved for both *Sinc* & Gaussian functions [31]. The Taylor expansion of the phase mismatched function around the perfectly phase matched frequency $\omega_\mu^0$ is as follows:

$$L\Delta k(\nu_s,\nu_i) = L\Delta k^0 + \tau_s\nu_s + \tau_i\nu_i + O(\nu^2) \quad (4)$$

The term $\Delta k^0$ represents perfectly phase matching condition ($\Delta k^0 \cong 0$) in the central frequency ($\omega_\mu^0$) and $\tau_\mu = L(u_\mu^{-1} - u_p^{-1})$ where $u_\mu$ ($\mu=i,s$) denotes the group velocity of the corresponding field [31]. Where the $\tau_\mu$ is known as temporal walk-off between generated photons and pump pulses. Since the amount of higher orders in the Eq. (4) which present the group velocity dispersion are negligible in comparison with the first order; in the following of the paper the first order of the Taylor expansion is considered. In addition the higher orders of the Eq. (4) also are three orders of magnitude less in comparison with the same effect in the external dispersive material. This is due to longer propagation length of the $l_\mu (\mu = s,i)$.

### III. Local detection

The typical method for illustrating time intervals between two photons is the HOM interferometer. Hong et al. [21] showed that if two indistinguishable photons arrive at the same time to a beam splitter (BS) they bunch together. Therefore, only one of the two detectors positioned in the output ports of the BS should be clicked. Hence the coincidence of two detectors is zero. This is in contrast with the situation where one of the photons delayed i.e. in its optical path. Therefore both detectors should be clicked simultaneously and their coincidence is not zero. This alteration in their phase (optical path) generates a HOM dip which its FWHM depends on the coherence length of the photons.

If a dispersive material like a fused silica (optical fiber) is placed between single photon source and the BS, then the effect of entanglement on the dispersion cancellation can be analyzed. Fig. 1b shows the HOM interferometer including a photon pair source followed by two external dispersive medium with $\beta_i$ and $\beta_s$ as their group velocity dispersion along propagation lengths of $l_i$ and $l_s$, respectively. An ultrashort pulse generates time-frequency entangled photons (Eq. (1)) through SPDC process. The generated entangled photons pass through dispersive material and mixed on the beam splitter (BS). A trombone system which is combination of an X-translation with two mirrors can be used for tuning of the relative phases between two photons via induced optical path difference. Finally the coincidence is measureable by two photodetectors.

The coincidence rate as a function of the delay time ($\tau$) induced in the pass length of one of the photons can be obtained by using correlation function of:

$$R_c(\tau) \propto \langle \psi_{SPDC} | E_1^-(t_1) E_2^-(t_2) E_1^+(t_1) E_2^+(t_2) | \psi_{SPDC} \rangle \quad (5)$$

Where $E_1$ and $E_2$ are the electric fields at each output ports of the BS and defined as follows:

$$E_1^+(t) = \frac{1}{\sqrt{2}}\left[E_s^+(t) + E_i^+(t-\tau)\right]$$
$$E_2^+(t) = \frac{1}{\sqrt{2}}\left[E_s^+(t) - E_i^+(t-\tau)\right] \quad (6)$$

In this equation $E_\mu^+(t) \propto \int d\omega_\mu\, a_\mu(\omega_\mu) e^{-i\omega_\mu t}$ and $\mu$ can be $s$ (signal) or $i$ (idler). Using Eq. (1-6) and after straight-forward calculations the coincidence rate of photon pair in local detection can be obtained as follows [29], [30]:

$$R_c^L(\tau) \propto 1 - \iint d\omega_s d\omega_i\, f(\omega_s,\omega_i) f^*(\omega_i,\omega_s) e^{i(\omega_s-\omega_i)\tau} \quad (7)$$

Where $f(\omega_s,\omega_i) = \alpha(\omega_s+\omega_i)\phi(\omega_s,\omega_i) D(\omega_s,\omega_i)$ is defined as modified joint spectral amplitude (JSA) function. The standard JSA is for the situation in which $D(\omega_s,\omega_i)=1$. Using the equations (1)-(6) and integrating the Eq. (7) it can be found that:

$$R_c^L(\tau) \propto 1 - \mathcal{P} \exp\left(\frac{-2\vartheta}{L^2\gamma(\tau_i-\tau_s)^2 \vartheta + 16\sigma_p^2(\beta_i^2-2\beta_i\beta_s+\beta_s^2)}\tau^2\right) \quad (8)$$

Where $\vartheta = 8 + L^2\gamma\sigma_p^2(\tau_i+\tau_s)^2$ and $\mathcal{P}$ demonstrates the visibility of the HOM interference. The coincidence rate as a function of interval arriving time ($\tau$) of two photon which is generated during SPDC process for an 810 nm pump with 4 nm FWHM of spectral bandwidth is illustrated in Fig. 2. A 10 mm long PPKTP is considered as the nonlinear crystal for the calculations presented in the rest of the paper. For a situation that there is no dispersive material in the photons pass lengths ($\beta_i = \beta_s = 0$) the calculated visibility is one and the HOM dip's width is 1.523 ps (solid red curve). However, when a dispersive material with $\beta_s = 100\times10^{-27}\,s^2$ (corresponding to propagation through a 20 m long fiber) is placed in front of the signal photon the width is broadened up to 1.726 ps (dashed blue curve) with decreased visibility down to $\mathcal{P}=0.88$.

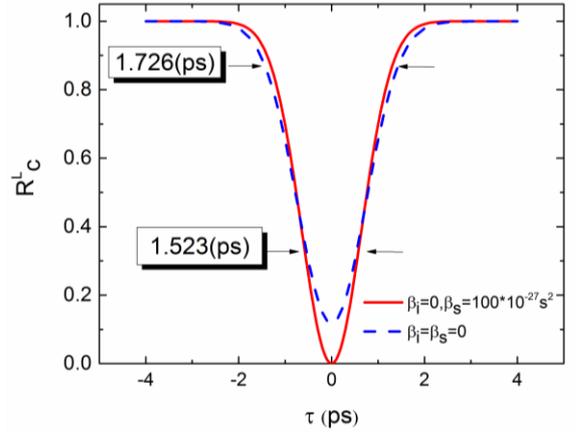

FIG. 2. Coincidence rate as a function of signal and idler's interval time. The solid red curve for a situation that there is no dispersive material and the dashed blue curve for a condition that a dispersive material with $\beta_s = 100\times10^{-27}\,s^2$ is in front of the signal photon.

The exponential function in Eq. (8) determines the FWHM of the HOM dip (local detection):

$$FWHM_L = \left(4\ln(2)\frac{L^2\gamma(\tau_i-\tau_s)^2 \vartheta + 16\sigma_p^2(\beta_i^2 - 2\beta_i\beta_s + \beta_s^2)}{2\vartheta}\right)^{1/2} \quad (9)$$

In the following of the paper this equation will be used to analyze the effect of different time-frequency modes on the broadening of HOM width caused by external dispersion. It is worth mentioning that, when the spectral bandwidth of the pump is small ($\sigma_p \to 0$), which indicates the continuous wave monochromatic pump source, the Eq. (9) is independent of the dispersive medium ($\beta_i$ & $\beta_s$):

$$FWHM_L = \sqrt{4\ln(2)}\left(\frac{L^2\gamma(\tau_i-\tau_s)^2}{2}\right)^{1/2} \quad (10)$$

The phenomenon which is known as dispersion cancellation and is reported by Steinberg and coworkers [10]. Also, if the generated entangled photons (signal & idler) propagate along dispersive materials with the same group velocity dispersion ($\beta_s = \beta_i$) the FWHM of HOM dip (Eq. (9)) is again independent of the dispersion (Eq. (10)). Also, Eq. (9) shows that by exchanging $\beta_i$ and $\beta_s$ the FWHM does not change. In other words if only one dispersive material is used, it is not important whether it is placed in idler arm or signal arm and FWHM is independent of the dispersive material place.

### IV. Non-local detection

The HOM interference is a local detection based measurement. However, the dispersion cancellation which is reported for the first time by Franson is based on a nonlocal detection [9] (see Fig. 1-a). The coincidence rate as a function of time delay can be calculated by Eq. (5) with the following electric field functions:

$$E_1^+(t) \propto \int d\omega_s \, a_s(\omega_s) e^{-i\omega_s t}$$
$$E_2^+(t) \propto \int d\omega_i \, a_i(\omega_i) e^{-i\omega_i(t-\tau)} \quad (11)$$

Like previous section, using Eq. (1-5) and Eq. (13) the FWHM of the coincidence rate of pair photons in a nonlocal detection can be obtained as follows:

$$FWHM_{NL} = \left(4Ln(2)\frac{8\beta_s^2 \mathcal{G}_i + \mathcal{Q}^2 + 8\beta_i^2 \mathcal{G}_s}{2\mathcal{Q}}\right)^{1/2} \quad (12)$$

Where $\mathcal{G}_\mu = 2 + L^2 \gamma \sigma_p^2 \tau_\mu^2$ and $\mathcal{Q} = L^2 \gamma (\tau_i - \tau_s)^2$. For a condition in which $\beta_i = \beta_s = 0$, the Eq. (12) is the same as Eq. (10). However, the main contrast between local (Eq.(9)) and nonlocal (Eq.(12)) detection is that whenever $\sigma_p \to 0$ the dispersion cancellation in nonlocal detection only occurs for the situation that $\beta_i = -\beta_s$. While for local detection FWHM of the HOM dip is independent of the dispersive element when $\sigma_p \to 0$. Moreover, it shows that the FWHM of the coincidence rate changes by exchanging $\beta_s$ and $\beta_i$. It means that if only one dispersive material is used it is important in which arm it is placed.

### V. Time Frequency Mode Engineering

Ultrashort pulsed lasers in SPDC process can be used to remove some uncertainty of the creation time of the generated single photon pair [32]. The process generates highly correlated time-frequency modes due to energy and momentum conservations law. The amount of this correlation can be found by applying a Schmidt decomposition to the JSA function [28]:

$$f(\omega_s, \omega_i) = \sum_j \sqrt{k_j} \, \varphi_j(\omega_s) \psi_j(\omega_i) \quad (13)$$

$\varphi(\omega_s)$ and $\psi(\omega_i)$ are two correlated orthonormal broadband frequency mode functions and $\sqrt{k_j}$ are the Schmidt coefficient and illustrates the amount of the correlations between time-frequency modes and satisfy $\sum_j k_j = 1$. The Schmidt number which shows the entanglement degree can be found as:

$$K = \frac{1}{\sum_j k_j^2} \quad (14)$$

The $K=1$ occurs in existence of a pure state with no entanglement of the generated state. As a consequent, the pure single photon state can be reached whenever the standard JSA function is separable and this is happened when it is oriented along one of the $\lambda_s$ or $\lambda_i$ coordinate axis [33].

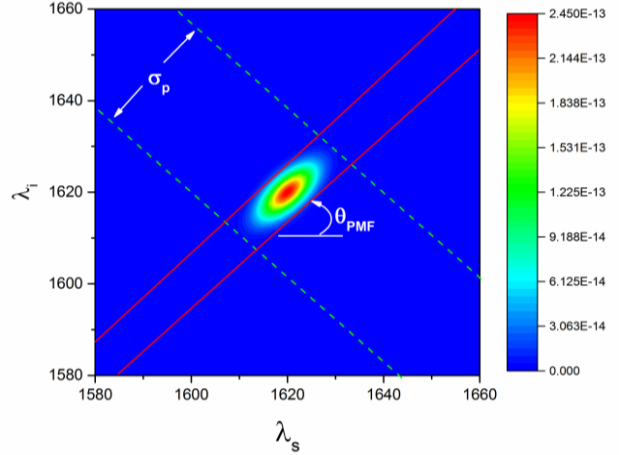

FIG. 3. The joint spectral amplitude in $\{\lambda_i, \lambda_s\}$-plane. The red line shows the phase matching orientation and the dotted green line shows the orientation of pump envelope. The value of $\sigma_p$ determines the width of pump envelope function.

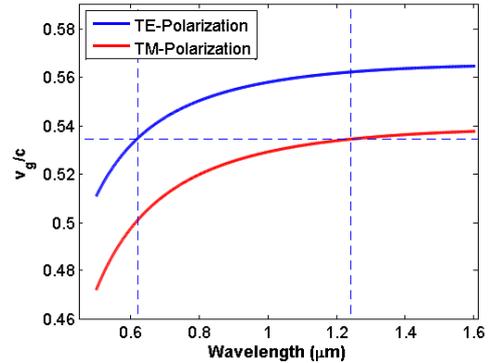

FIG. 4. Variation of the group velocity as a function of the wavelength for a PPKTP. The blue line is the TE-polarization (Pump and Signal) and red line shows TM-Polarization (Idler) in a type II SPDC configuration. The dotted lines illustrate the wavelength (620nm) in which the group velocity of the pump is equal to the idler's (1240nm) group velocity.

The JSA orientation depends on the angle of the phase matching function ($\phi(\omega_s, \omega_i)$) and the FWHM of the pump spectral bandwidth which is introduced in Eq. (1) in the $\{\lambda_S, \lambda_i\}$-plane. In Fig. 3 a typical JSA is illustrated for a 10 mm long PPKTP crystal pumped by pulsed 810 nm with its FWHM of 4 nm. According to the conservation of momentum ($\Delta k = 0$) this angle can be obtained from Eq. (4):

$$\Theta_{PMF} = -\arctan\left(\frac{\tau_s}{\tau_i}\right) \quad (15)$$

The separation of JSA function is reported experimentally via crystal dispersion engineering in which the wavelength of the input pump tuned to have same group velocity as the signal/idler photon [27]. In the other words, if the $\tau_s = 0$ ($\tau_i = 0$) the JSA will be parallel along $\lambda_S$-plane with $\Theta_{PMF} = 0$ ($\lambda_i$-plane with $\Theta_{PMF} = \pm\frac{\pi}{2}$). The variations of the group velocities of the PPKTP in both ordinary (TE-Polarization) and extraordinary directions (TM-Polarization) as a function of wavelength are illustrated in Fig.4. In the type II SPDC configuration, input pump has horizontal polarization parallel to the ordinary refractive index. Therefore, the pump wavelength in which its group velocity is the same as the idler is 620 nm.

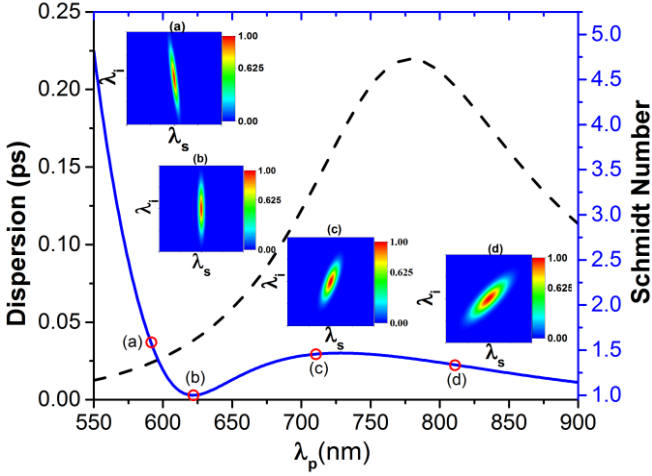

FIG. 5. Right axis (solid blue curve) shows the Schmidt number vs pump wavelength. Left axis (dashed black curve) shows the difference FWHM of HOM dip between $\beta_s = 100 \times 10^{-27} s^2$ and $\beta_s = 0$. The $\beta_i = 0$ is assumed to be zero. Inset shows the JSA function in $\{\lambda_i, \lambda_s\}$-plane for 590nm (a), 620nm (b), 710nm (c) and 810nm (d) pump wavelengths.

For this wavelength (620 nm) the phase matching function will be parallel with $\lambda_i$ axis. So, its correlation with a pump envelope with $\sigma_p = 4$ nm causes a JSA parallel with $\lambda_i$ axis as it is shown in the inset-a of the Fig. 5. In this situation Schmidt number $K=1$ and, therefore, the generated photon pair are not entangled. In the Fig. 5 the calculated Schmidt number as a function of pump wavelength is illustrated (solid blue curve). The JSA for 590 nm, 620 nm, 710 nm and 810 nm pump wavelengths are also shown as insets-a, -b, -c and -d respectively. According to the Eq. (15) and Eq. (14), for different wavelengths various degrees of entanglement are achievable except for 620nm wavelength where the Schmidt number is equal to one.

## VI. Time Frequency Mode Impact on Dispersion

In the following of the paper the discussion is concentrated on the collinear type II phase matching (daughter photons are orthogonally polarized) in a PPKTP crystal. It is worth mentioning that the achieved results are independent of the used nonlinear medium and it can be concluded to all other nonlinear crystals which are used for generating heralded single photons.

Fig. 5 (dashed black curve) presents the difference between the FWHM of HOM dip for $\beta_s = 100 \times 10^{-27} s^2$ and $\beta_s = 0$.

The spectral bandwidth of $\sigma_p = 4$ nm is considered in Eq. (10) with $\beta_s = 0$. Since for pump wavelength of 620 nm group velocity difference between pump and idler is zero, the generated state is not entangled. Therefore, it is expected that in the presence of an external dispersive material the maximum broadening should be occurred. This is in contrast with what is presented in Fig. 5 (dashed black curve) that the maximum broadening around 800 nm is observed.

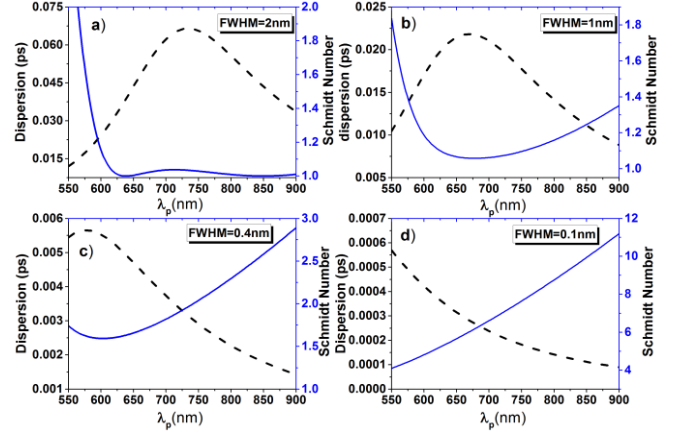

FIG. 6. The variation of Schmidt number vs pump wavelength (solid blue curves) and difference FWHM of HOM dip between $\beta_s = 100 \times 10^{-27} s^2$ and $\beta_s = 0$ as a function of pump wavelength (dashed black curves). Different curves present the pump spectral bandwidths of 2 nm (a), 1 nm (b), 0.4 nm (c) and 0.1 nm (d).

By minimizing of the denominator of the fraction in the Eq. (10) ($\tau_i + \tau_s = 0$) FWHM$_{HOM}$ is maximum. In the other words whenever the $\Theta_{PMF}$ is 45 degree with respect to $\{\lambda_S, \lambda_i\}$-plane the maximum broadening happened. In this situation the temporal walk-off between signal and pump is the same as idler's temporal walk-off but with opposite sign. Again this reveals that the broadening of the dispersion is related to the group velocity of generated signal and idler photons and independent of their entanglement for the pump with 4 nm spectral bandwidth. In the other words, it shows that time-frequency modes have no effect on the dispersive broadening for such a spectral bandwidth.

In Fig. 6 the same investigation is done on different pump spectral bandwidth. For a pump with FWHM=2 nm (Fig. 6a) the pick of the dispersive broadening moves towards shorter wavelengths where the degree of the entanglement decreases. It illustrates that by decreasing the pump spectral bandwidth the entanglement has more effect on the compensation of dispersive broadening. This phenomenon is obvious for a pump with 1 nm (Fig. 6b) and 0.4 nm (Fig. 6c) spectral bandwidth. By manipulating the number of time frequency modes (decreasing the entanglement degree) the dispersion compensation decreases, too. Fig. 6d clearly shows that Schmidt number and dispersive broadening are in the opposite manner. By increasing the pump wavelength the Schmidt number or in other words entanglement degree increases and dispersive broadening decreases.

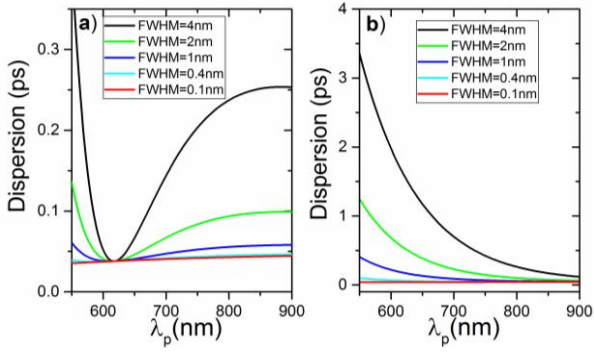

FIG. 7. Representation of dispersion as a function of pump wavelength for different FWHM of coincidence rate between $\beta_s = 100 \times 10^{-27} s^2$ and $\beta_s = 0$ when the input pump has the spectral bandwidth of 4 nm (black) 2 nm (green), 1 nm (blue), 0.4 nm (magenta) and 0.1 nm (red). a) The dispersive material is in front of the signal photon. b) The dispersive material is in front of idler photon.

In non-local detection a completely different results are obtained when the effect entanglement is considered. In Fig. 7 the difference FWHM of coincidence rate between $\beta_s = 100 \times 10^{-27} s^2$ and $\beta_s = 0$ as a function of different pump wavelength is illustrated. It shows that the FWHM of the coincidence rate depends on the arm which dispersive material is placed. When the dispersive material is in the idler arm the FWHM is uniform but for a situation that the dispersive material is in the signal arm the FWHM is not uniform. The maximum compensation is for a situation in which $\Theta_{PMF} = \dfrac{\pi}{2}$. It means the entanglement does not have any influence on dispersive broadening compensation when only one dispersive material is used in nonlocal detection.

## VII. Conclusion

Dispersion cancelation in both nonlocal and local detections is reported with the argument of its originality from consequence of entangled state preparation of the interacting pairs of photons [15], [17]. On the other hand, the entanglement state engineering of the photon pair can be achieved via pump spectral manipulation in a type II SPDC process [1], [27]. Such a state engineering is applied to the dispersion cancelation in both scenarios of local and nonlocal detections to examine the impact of entanglement of involved pair of photons. The Schmidt mode numbers for SPDC-based generated daughter photons in PPKTP crystal (for instance) are calculated revealing that tuning the wavelength of the pulsed input pump would result to manipulate the entanglement degree of the generated photon pair. For the pump wavelength of 620 nm with 4 nm spectral bandwidth the Schmidt number is equal to one indicating no entanglement for the generated pair of photons. Based on similar calculations, the maximum entanglement can be achieved in 800 nm of the input pump wavelength. According to the results presented in Ref. [Steinberg] the maximum dispersion cancelation also should be observed at this wavelength, whereas, the obtained results illustrate the maximum dispersion cancelation in different wavelength. This indicates that the entanglement does not have influence on the dispersive broadening when the pump spectral bandwidth is large (here 4 nm). In сontrast to the entanglement effect, our results for local detection illustrate the strong impact of temporal walk-offs between signal/idler and pump in the broadening effect. However, it is illustrated that by decreasing the pump spectral bandwidth the entanglement degree has the major effect on the dispersive broadening compensation.

The same investigation is performed for nonlocal detection. In this case, the broadening depends on the position of the dispersive material in such a way that the broadening is different when the dispersive material is in signal arm or in the idler arm.